\documentclass{JHEP3}
\usepackage{amssymb}

\def \be {\begin{equation}}
\def \ee {\end{equation}}
\newcommand{\eq}[1]{(\ref{#1})}
\def \bea {\begin{eqnarray}}
\def \eea {\end{eqnarray}}
\def \nn {\nonumber}

\def \a {\alpha}
\def \b {\beta}

\def \G {\Gamma}
\def \d {\delta}

\def \m {\mu}
\def \n {\nu}
\def \k {\kappa}

\def \s {\sigma}
\def \r {\rho}
\def \o {\omega}
\def \O {\Omega}
\def \th {\theta}
\def \Th {\Theta}

\def \t {\tau}
\def \dag {\dagger}
\def \p {\partial}

\def\bd{\begin{document}}
\def\ed{\end{document}}
\def\nn{\nonumber}
\def\bea{\begin{eqnarray}}
\def\eea{\end{eqnarray}}
\let\bm=\bibitem
\let\la=\label

\def\N{{\cal N}}
\def\sst{\scriptscriptstyle}
\def\thetabar{\bar\theta}
\def\Tr{{\rm Tr}}
\def\one{\mbox{1 \kern-.59em {\rm l}}}

\def\a{\alpha}      \def\da{{\dot\alpha}}
\def\b{\beta}       \def\db{{\dot\beta}}
\def\c{\gamma}  \def\C{\Gamma}  \def\cdt{\dot\gamma}
\def\d{\delta}  \def\D{\Delta}  \def\ddt{\dot\delta}
\def\e{\epsilon}        \def\vare{\varepsilon}
\def\f{\phi}    \def\F{\Phi}    \def\vvf{\f}
\def\h{\eta}
\def\k{\kappa}
\def\l{\lambda} \def\L{\Lambda}
\def\m{\mu} \def\n{\nu}
\def\o{\omega}
\def\P{\Pi}
\def\r{\rho}
\def\s{\sigma}  \def\S{\Sigma}
\def\t{\tau}
\def\th{\theta} \def\Th{\Theta} \def\vth{\vartheta}
\def\X{\Xeta}
\def\z{\zeta}
\def\w{\wedge}
\def\u{\underline}
\def\hs{\hspace}

\def\cA{{\cal A}} \def\cB{{\cal B}} \def\cC{{\cal C}}
\def\cD{{\cal D}} \def\cE{{\cal E}} \def\cF{{\cal F}}
\def\cG{{\cal G}} \def\cH{{\cal H}} \def\cI{{\cal I}}
\def\cJ{{\cal J}} \def\cK{{\cal K}} \def\cL{{\cal L}}
\def\cM{{\cal M}} \def\cN{{\cal N}} \def\cO{{\cal O}}
\def\cP{{\cal P}} \def\cQ{{\cal Q}} \def\cR{{\cal R}}
\def\cS{{\cal S}} \def\cT{{\cal T}} \def\cU{{\cal U}}
\def\cV{{\cal V}} \def\cW{{\cal W}} \def\cX{{\cal X}}
\def\cY{{\cal Y}} \def\cZ{{\cal Z}}

\def\ua{\underline{\alpha}} \def\ubb{\underline{\beta}}
\def\ug{\underline{\gamma}}
\def\ub{\underline{\phantom{\alpha}}\!\!\!\beta}
\def\uc{\underline{\phantom{\alpha}}\!\!\!\gamma}
\def\um{\underline{\mu}} \def\un{\underline{\nu}}
\def\ud{\underline\delta}
\def\ue{\underline\epsilon}
\def\una{\underline a}\def\unA{\underline A}
\def\unb{\underline b}\def\unB{\underline B}
\def\unc{\underline c}\def\unC{\underline C}
\def\und{\underline d}\def\unD{\underline D}
\def\une{\underline e}\def\unE{\underline E}
\def\unf{\underline{\phantom{e}}\!\!\!\! f}\def\unF{\underline F}
\def\unm{\underline m}\def\unM{\underline M}
\def\unn{\underline n}\def\unN{\underline N}
\def\unp{\underline{\phantom{a}}\!\!\! p}\def\unP{\underline P}
\def\unq{\underline{\phantom{a}}\!\!\! q}
\def\unQ{\underline{\phantom{A}}\!\!\!\! Q}
\def\unH{\underline{H}}
\def\ul{\underline}

\def\As {{A \hspace{-6.4pt} \slash}\;}
\def\bs {{b \hspace{-6.4pt} \slash}\;}
\def\Ds {{D \hspace{-6.4pt} \slash}\;}
\def\ds {{\del \hspace{-6.4pt} \slash}\;}
\def\ss {{\s \hspace{-6.4pt} \slash}\;}
\def\ks {{ k \hspace{-6.4pt} \slash}\;}
\def\ps {{p \hspace{-6.4pt} \slash}\;}
\def\pas {{{p_1} \hspace{-6.4pt} \slash}\;}
\def\pbs {{{p_2} \hspace{-6.4pt} \slash}\;}

\def\Fh{\hat{F}}
\def\Vh{\hat{V}}
\def\Xh{\hat{X}}
\def\ah{\hat{a}}
\def\xh{\hat{x}}
\def\yh{\hat{y}}
\def\ph{\hat{p}}
\def\xih{\hat{\xi}}

\def\psit{\tilde{\psi}}
\def\Psit{\tilde{\Psi}}
\def\tht{\tilde{\th}}

\def\At{\tilde{A}}
\def\Qt{\tilde{Q}}
\def\Rt{\tilde{R}}
\def\Nt{\tilde{N}}

\def\at{\tilde{a}}
\def\st{\tilde{s}}
\def\ft{\tilde{f}}
\def\pt{\tilde{p}}
\def\qt{\tilde{q}}
\def\vt{\tilde{v}}
\def\nt{\tilde{n}}

\def\delb{\bar{\partial}}
\def\bz{\bar{z}}
\def\bD{\bar{D}}
\def\bB{\bar{B}}
\def\bo {\bar{\o}}

\def\bk{{\bf k}}
\def\bl{{\bf l}}
\def\bp{{\bf p}}
\def\bq{{\bf q}}
\def\br{{\bf r}}
\def\bx{{\bf x}}
\def\by{{\bf y}}
\def\bR{{\bf R}}
\def\bV{{\bf V}}
\def\bd{\begin{document}}

\def\ed{\end{document}}

\def\d{\delta}\def\D{\Delta}\def\ddt{\dot\delta}

\def\p{\partial} \def\del{\partial}
\def\xx{\times}
\def\uno{\mbox{1 \kern-.59em {\rm l}}}

\def\trp{^{\top}}
\def\inv{^{-1}}
\def\dag{{^{\dagger}}}
\def\pr{\prime}

\def\rar{\rightarrow}
\def\lar{\leftarrow}
\def\lrar{\leftrightarrow}

\def\cw{{\cal W}}
\def\cz{{\cal Z}}
\def\tcm{\tilde{\cal M}}
\def\sgn{{\rm sgn}}
\def\sd {d^{4|4}}
\def\lan{\langle}
\def\ran{\rangle}

\title{Real-time correlators in Kerr/CFT correspondence}
\author{Bin Chen\\
Department of Physics,\\
and State Key Laboratory of Nuclear Physics and Technology,\\
and Center for High Energy Physics,\\
Peking University,\\
Beijing 100871, P.R. China\\
\email{bchen01@pku.edu.cn}}

\author{Chong-Sun Chu\\
Centre for Particle Theory and Department of Mathematics,\\
Durham University,\\
Durham, DH1 3LE, UK\\
\email{chong-sun.chu@durham.ac.uk}}

\date{\today}

\abstract{We study real-time correlators in the Kerr/CFT
correspondence. The near-horizon geometry of extreme
and near-extremal Kerr black holes could be taken as the
warped AdS spacetimes with the warping factor being a function
of angular variable $\th$. We show that for the perturbations
whose equations of motions could be decomposed into the angular
part and radial part, their real-time correlation functions could
be computed from warped AdS/CFT correspondence. We find that the retarded
Green's functions, the cross sections and the quasi-normal modes are all
in perfect match with the dual CFT predictions. The same analysis
is also generalized to the charged Newman-Kerr black holes.}

\preprint{DCPT-10/01}

\newpage
\begin{document}

\section{Introduction}\label{sec-intro}

Much of the physics on black hole is encoded in the near-horizon
geometry of the black hole. For example, the entropy of the black
hole is just proportional to the area of the black hole horizon.
And the Hawking radiation of the black hole could be effectively
understood from the anomaly cancellation of the field theory in
the near horizon geometry \cite{Robinson:2005pd,{Xu:2006tq}}.
A recent evidence to this universal
property is the Kerr/CFT correspondence  \cite{AndyWei}.

The Kerr/CFT  correspondence conjectures
that the quantum gravity in the near-horizon extreme Kerr
(NHEK) geometry with certain
boundary conditions is dual to a (1+1) dimensional chiral conformal
field theory (CFT)
\footnote{More precisely, the Kerr/CFT correspondence can be referred
 to as the NHEK/CFT correspondence.}.
The correspondence
was inspired by the  properties of the
asymptotic symmetry group of the near horizon geometry \cite{Brown86}
of the extreme
Kerr black hole where it was found
by Guica, Hartman, Song and Strominger (GHSS)
\cite{AndyWei} that under a
certain  set of  boundary condition on the asymptotic behaviour of the
metric, the $U(1)_L$ symmetry of the  $SL(2,R)_R \times U(1)_L$
isometry group \cite{Bardeen:1999px} of the
near-horizon
geometry get enhanced into a Virasoro algebra.
Support
of this conjecture has been found in  the perfect match of the macroscopic
Berenstein-Hawking entropy of the  black hole with the conformal field
theory entropy computed by the  Cardy formula. See  \cite{Lu:2008jk}
for some further studies  of the Kerr/CFT correspondence as well as
generalizations to other spacetime which contain a warped AdS
structure.

Further support of the correspondence are found in the studies of
the superradiant scattering processes
off extreme Kerr black holes
\cite{Bredberg:2009pv}.
In this case, the near horizon geometry
of a near-extremal Kerr black hole
(near-NHEK) is reminiscent of a
non-extremal warped black hole. Correspondingly, the right-moving
sector of dual CFT is excited \cite{Castro:2009jf}.
An important ingredient in the studies
is the Teukolsky master equations \cite{Teukolsky:1973ha,Press:1973zz,
  {Teukolsky:1974yv},{Starobinsky1},Starobinsky2}, which include an
angular equation and a radial equation. In the near-horizon limit, the
modes of interest are the ones near the super-radiant bound. This
implies that the separation constant is well independent of the
frequency of the mode and so the radial equation can be decoupled
completely from the angular equation. This allows
\cite{Bredberg:2009pv} to compute the quantum decay rate of the bulk
fields and
the greybody factor of the extreme Kerr black hole. On the CFT side,
the decay rate and the absorption cross section can be extracted from
the two-point correlation function \cite{Maldacena:1997ih}.
It is remarkable that the bulk scattering results
are in precise agreement with the CFT description whose form is
completely fixed by conformal invariance.
Similar discussion has been generalized to charged
Kerr-Newman \cite{Hartman:2009nz}, multi-charged
Kerr \cite{Cvetic:2009jn} and higher dimensional near-extremal Kerr
black holes. In all these cases, perfect agreement with the dual CFT
description has been found.

Note that in these tests of the Kerr/CFT correspondence,
the fundamental CFT two-point correlator is compared with
secondary quantities such as the decay rates
derived from the superradiant scattering processes.
We recall that in the standard AdS/CFT correspondence,
it is possible to extract the CFT real-time correlator directly from
the bulk asymptotic AdS spacetime \cite{Son05}.  This prescription has also
been shown to work \cite{Chen:2009cg}
for the warped AdS/CFT correspondence \cite{Andy08} as well
\footnote{For a discussion
of other related  issues of the warped AdS/CFT correspondence, see for
example  \cite{{Compere:2008cv},{Compere:2009zj}, {Anninos:2009zi},
{Blagojevic:2009ek},{ChenXu09},{ChenXu2}}. }.
It is natural to ask if one can also compute the
real-time correlators directly from the bulk side for
the Kerr/CFT correspondence. This will allow one to perform a test
directly on the CFT correlators and the real-time correlators as obtained by
holography. Now although the NHEK geometry is more complicated, it is
in fact a warped AdS${}_3$  spacetime with a
warping factor being a function of the angular variable, therefore one can
consider the  Kerr/CFT correspondence as a generalization
of the warped AdS/CFT correspondence.  In this paper, we show that with
a small modification, the  Minkowskian prescription for computing the real-time
correlators continues to work.
The results are in perfect agreement
with the CFT predictions.

In the next section, we give a brief review of Kerr/CFT
correspondence for extreme and near-extremal Kerr black holes.
In section 3, we
outline the forms of the correlation functions as determined by the
conformal invariance in 2D conformal field theory. In
section 4 and 5, by considering the  Kerr/CFT
correspondence as a generalized warped AdS/CFT correspondence, we
use the Minkowskian prescription   to compute retarded correlators of various
fields in near-NHEK and NHEK case respectively. We will
show that the results agree precisely with the CFT correlators
provided that one absorb away
an angular-dependent multiplicative factor due to
the
angular dependent
warping factor.
In section 6, we discuss the charged Kerr-Newman case.
We end with some discussions in section 7.

\section{Brief Review of the Kerr/CFT Correspondence}

A Kerr black hole is characterized by the mass $M$ and angular
momentum $J=aM$. It could be described by the metric of the following
form
 \be
 ds^2=-\frac{\D}{\hat{\rho}^2}(d\hat t-a\sin^2\th d\hat
 \phi)^2+\frac{\sin^2\th}{\hat \rho^2}\left((\hat r^2+a^2)d\hat
 \phi -a d\hat t\right)^2+\frac{\hat \rho^2}{\D}d\hat r^2+\hat
 \rho^2 d\th^2,
 \ee
 with
 \be
 \D=\hat r^2-2M\hat r +a^2, \hs{5ex} \hat \rho^2=\hat r^2+a^2
 \cos^2 \th,
 \ee
where we have used the unit $G= \hbar =c =1$.
 In general, there are two horizons at $\D=0$, which gives
 \be
 r_\pm = M\pm \sqrt{M^2-a^2}.
 \ee
 The Hawking temperature,
the angular velocity of the horizon
and the entropy of the Kerr black hole
 are
 \be
 T_H=\frac{r_+-r_-}{8\pi Mr_+}, \hs{5ex}
\Omega_H = \frac{a}{2M r_+},  \hs{5ex}
S_{BH}=2\pi M r_+.
 \ee
 For the extreme Kerr black hole, the Hawking temperature is
 zero, and its entropy is
 \be
 S_{\rm ext}=2\pi J=2\pi M^2.
 \ee

We are interested in the extreme and near-extremal Kerr black
holes. First we focus on the near-extremal Kerr case, whose near
horizon geometry could be defined by taking the limit $T_H \to 0$
and $\hat r\to r_+$
and with the
dimensionless
near-horizon temperature
\be T_R
\equiv \frac{2M T_H}{\l}
\ee
being fixed when $\l \to 0$. In other
words, even though the Hawking temperature at  asymptotic infinity
is zero, the temperature measured near the horizon is still finite
due to the infinite blueshift. For the extreme black hole, $T_R$
is exactly zero even at the horizon.
As in  \cite{Bredberg:2009pv}, for the near-extremal Kerr, we have
 \bea
 r_+&=& M+\l M 2\pi T_R +O(\l^2), \\
 a&=& M-2M(\l \pi T_R)^2+O(\l^3).
 \eea
 After redefining the coordinates
 \bea
 t&=& \l \frac{\hat t}{2M}, \\
 r&=&\frac{\hat r-r_+}{\l r_+}, \\
 \phi&=&\hat \phi - \frac{\hat t}{2M},
 \eea
 and keeping $T_R$ fixed, we
obtain the near-extremal near-horizon
 metric
 \be\label{near-NHEK}
 ds^2=2J\G\left(-r(r+2\a)dt^2 + \frac{dr^2}{r(r+2\a)}+d\th^2
+\L^2(d\phi+(r+\a)dt)^2\right),
 \ee
 where $\a=2\pi T_R$, \be
 \G(\th)=\frac{1+\cos^2\th}{2},
 \hs{3ex}\L(\th)=\frac{2\sin\th}{1+\cos^2\th}
 \ee
and $\phi \sim \phi +2\pi, 0\leq \th \leq \pi$.
For the extreme Kerr black hole, $T_R=0$, its near-horizon
geometry in Poincare-type coordinates is
\cite{Bardeen:1999px}
 \be\label{NHEK}
 ds^2=2J\G\left(-r^2dt^2 +
 \frac{dr^2}{r^2}+d\th^2+\L^2(d\phi+rdt)^2\right).
 \ee
  In global
coordinates, the metric is
 \be\label{global}
ds^2=2J\G\left(-(1+\rho^2)d\t^2 +
\frac{d\rho^2}{1+\rho^2}+d\th^2+\L^2(d\varphi+\rho d\t)^2\right).
 \ee

The NHEK geometry has an isometry $SL(2,R)_R \times U(1)_L$. Moreover,
for each three-dimensional slice of fixed polar angle
$\th$, (\ref{global}) is a global warped AdS${}_3$ spacetime, while
the NHEK geometry (\ref{NHEK}) is the quotient of warped AdS${}_3$.
In other words,
just as the BTZ
black hole as the quotient of AdS${}_3$ \cite{BTZ},
the global warped AdS spacetime \eq{global} can be taken as the
vacuum with the geometry \eq{NHEK} (resp. \eq{near-NHEK}) being taken
as an extreme warped AdS${}_3$ black hole
(resp.
as an non-extremal warped AdS${}_3$ black hole).
Note that although the metrics
(\ref{NHEK}) and (\ref{global}) can be related by the coordinate
transformation, but they describe different geometries since the
coordinate transformation is singular.

It was shown in  \cite{AndyWei} that with consistent boundary
condition the $SL(2,R)_R$ becomes trivial while the $U(1)_L$ is
enhanced to a Virasoro algebra with central charge
$c_L= 12J$. Moreover, the quantum theory in the
Frolov-Thorne vacuum for extreme Kerr has the left-moving
temperature $T_L=\frac{1}{2\pi}$.
Therefore the Cardy formula gives the entropy for the dual CFT
 \be
 S=\frac{\pi^2}{3}c_L T_L = 2\pi J .
 \ee
This matches exactly with the Bekenstein-Hawking
entropy of the extreme Kerr black hole.
It is important to emphasize that the Kerr/CFT correspondence is
a correspondence between the quantum gravity in the NHEK
geometry and a  2D chiral CFT with $c_L$ and $T_L$.
In fact as shown in
 \cite{Bredberg:2009pv}, the CFT cross section only counts for the
scattering amplitude near the horizon region rather than the whole
extreme Kerr black hole including the asymptotic region.

For the near-extremal Kerr black hole, its
near horizon geometry is
(\ref{near-NHEK}).
The entropy of the near-extremal Kerr black hole is
the  same as
for the extreme one.
Now in
dual 2D CFT, the right-moving sector is excited with a finite
temperature. The Cardy formula gives the entropy
 \be
 S=\frac{\pi^2}{3}(c_L T_L+c_R T_R).
 \ee
In order to match with the black hole entropy,
the central charge in the right-moving sector should
be zero.
It will be interesting to find the consistent boundary conditions
that lead to an asymptotic symmetry group which extends the
$SL(2,R)_R$ to Virasoro and compute the central charge to confirm
this.


\section{Two-point Correlators in 2D CFT}

In a 2D conformal field theory(CFT), one can define the two-point function
\be
 G(t^+,t^-)=\langle {\cal O}^\dagger_\phi(t^+,t^-){\cal
 O}_\phi(0)\rangle,
\ee
where $t^+,t^-$ are the left and right moving coordinates of 2D
worldsheet and  ${\cal O}_\phi$ is the operator corresponding to
the field perturbing the black hole.
For an operator of  dimensions $(h_L,h_R)$,
charges $(q_L,q_R)$ at temperature $(T_L,T_R)$ and chemical potentials
$(\O_L, \O_R)$, the two-point function  is
dictated by conformal invariance and takes the form
\cite{Cardy:1984bb}:
 \be \label{G-Mink}
 G(t^+,t^-)\sim (-1)^{h_L+h_R}\large(\frac{\pi T_L}{\sinh(\pi T_L
 t^+)}\large)^{2h_L}\large(\frac{\pi T_R}{\sinh(\pi T_R
 t^-)}\large)^{2h_R}e^{iq_L\O_L t^+ +iq_R\O_R t^-}.
 \ee
In frequency domain, the  decay rate  and the
CFT absorption cross section are given by
\cite{Maldacena:1997ih}
\bea
\G &\sim& \int dt^+ dt^- e^{-i \o_R t^- -i \o_L t^+} G(t^+-i \e,t^- -i \e),\nn\\
\s_{\rm abs} &\sim&  \int dt^+ dt^- e^{-i \o_R t^- -i \o_L t^+}
[ G(t^+-i \e,t^- -i \e) - G(t^+ + i \e,t^- +i \e)] .
\eea
It follows  that
\bea
\G &\sim&  T_L^{2 h_L-1}  T_R^{2 h_R-1}
e^{- \bo_{L} / 2T_L} e^{ - \bo_{R}/ 2T_R}
|\G(h_L + i\frac{\bo_{L}}{2 \pi T_L})|^2
|\G(h_R + i\frac{\bo_{R}}{2 \pi T_R})|^2, \label{g1}
\eea
and $ \s_{\rm abs} \sim \G(\bo_L,\bo_R) \pm\G(-\bo_L,-\bo_R)$ giving
\bea \label{g2}
\s_{{\rm abs}} \sim
T_L^{2 h_L-1}  T_R^{2 h_R-1}|\G(h_L + \frac{i\bo_{L}}{2 \pi T_L})|^2
|\G(h_R + \frac{i\bo_{R}}{2 \pi T_R})|^2 \cdot \bigg\{
\begin{array}{cc}
\sinh(\frac{\bo_{L}}{2T_L} + \frac{\bo_{R}}{2T_R}),& \mbox{\small (boson)}\\
\cosh(\frac{\bo_{L}}{2T_L} + \frac{\bo_{R}}{2T_R}).&
\mbox{\small (fermion)}\end{array} \;\;\;\;\;
\eea
Here $\bo_L, \bo_R$ are defined by
\be
\bo_L= \o_L - q_L \O_L, \quad \bo_R= \o_R - q_R \O_R.
\ee
The $\sim$ sign here (and below)
means the LHS and the RHS are equal up to a factor
independent of the frequencies.
A perfect match has been found for the expressions \eq{g1}, \eq{g2}
with the macroscopic decay rate and the greybody factor for a bulk
field scattering off (near)-NHEK \cite{Bredberg:2009pv}. This gives
strong support to the Kerr/CFT correspondence.

Apart from the relations \eq{g1}, \eq{g2},
another connection of the two-point function with the bulk
can be developed.
Let  us introduce the Euclidean correlator $G_E$
by a Wick rotation $t^+ \to i \t_L$, $t^- \to i \t_R$.
At finite
temperature the Euclidean time is taken to have period
$2 \pi/T_L, 2 \pi/T_R$ and  the momentum space Euclidean correlator
is given by
\be
G_E(\o_{L,E}, \o_{R,E}) = \int_{0}^{2\pi/T_L} d\t_L \int_{0}^{2\pi/T_R} d \t_R\;
e^{-i \o_{L,E}\t_L - i\o_{R,E}\t_R} G_E(\t_L,\t_R),
\ee
where the Euclidean frequencies are related to the Minkowskian ones by
\be
\o_{L,E} = i \o_L, \qquad \o_{R,E} = i \o_R.
\ee
The integral is divergent but can be defined by analytic
continuation, one obtains \cite{Maldacena:1997ih}
\bea \label{GE}
G_E(\o_{L,E}, \o_{R,E})  \sim T_L^{2 h_L-1}  T_R^{2 h_R-1}
e^{i \frac{\bo_{L,E}}{2T_L}} e^{i \frac{\bo_{L,E}}{2T_R}}
\G(h_L + \frac{\bo_{L,E}}{2 \pi T_L})\G(h_L - \frac{\bo_{L,E}}{2 \pi T_L})
\nn\\
\cdot\G(h_R + \frac{\bo_{R,E}}{2 \pi T_R})\G(h_R - \frac{\bo_{R,E}}{2 \pi T_R}),
\eea
where
\be
\bo_{L,E}= \o_{L,E} - i q_L \O_L, \quad \bo_{R,E}= \o_{R,E} - i q_R \O_R.
\ee

We note that $G_E(\o_{L,E}, \o_{R,E})$ is related to the value of the
retarded correlator $G_R (\o_L, \o_R)$. More specifically, the retarded
correlator  $G_R (\o_L, \o_R)$ is analytic on the upper half complex
$\o_{L,R}$-plane and its value along the positive
imaginary $\o_{L,R}$-axis
gives the Euclidean correlator:
\be \label{GER}
G_E(\o_{L,E}, \o_{R,E}) = G_R(i\o_{L,E}, i\o_{R,E}), \quad \o_{L,E} , \o_{R,E} >0.
\ee
This relation holds both for zero and finite temperature. However
at finite temperature,
$\o_{L,E}$ and $\o_{R,E}$ take discrete values
of the Matsubara frequencies

\be
\o_{L,E} =  2 \pi m_L T_L, \quad \o_{R,E} =  2 \pi m_R T_R,
\ee
where $m_L, m_R$ are integers for bosonic modes and are half integers
for fermionic modes.
As we mentioned in the introduction, since the retarded
correlator can be computed directly
with a holographic prescription in terms of the bulk, therefore \eq{GER}
provides a direct test of the Kerr/CFT correspondence.

\section{Real-time Correlators in Near-NHEK of Kerr Black Hole}

In the AdS/CFT correspondence, one subtle point in the computation of
real-time correlators is the boundary conditions of the classical
solution at the black hole horizon. Different Green's functions
correspond to different boundary conditions at the horizon. It
turns out that the retarded Green's function corresponds to the
ingoing boundary condition, while the advanced Green's function
corresponds to the outgoing one. However, even after fixing the
boundary condition, one cannot obtain the correlators by
naively using the prescription in the Euclidean version of the AdS/CFT
correspondence.

In  \cite{Son05}, a simple prescription was
proposed to compute real-time correlators from gravity. This
prescription has been instrumental to the study of strongly
interacting system at finite temperature during the past few
years. It has also been justified from different points of view in
 \cite{Herzog:2002pc,Marolf:2004fy,Gubser:2008sz,
Skenderis:2008dg,Iqbal:2008by,Iqbal:2009fd}.
It was later observed \cite{Gubser:2008sz} that
this prescription could be recast in terms of the boundary values of
the canonical conjugate momentum of the bulk fields by treating
the AdS radial direction as ``time" direction.
Furthermore, this reformulation was shown to be able to follow
directly from the analytic continuation of Euclidean AdS/CFT
correspondence \cite{Iqbal:2009fd}. More precisely, for a background metric
\be
ds^2 = g_{rr} dr^2 + g_{\m\n} dx^\m dx^\n,
\ee
where $\m,\n$ run over a $d$-dimensional spacetime, assume that the metric
has an event horizon at $r=r_0$ and a boundary at $ r= \infty$. Also assume
that all metric components depends only on $r$, then
the prescription for computing the retarded correlator is:
 \be\label{GR}
 G_R(\o, \vec{k})\sim
\bigg(
\lim_{r\to \infty} r^N \frac{\Pi(r,\o,
 \vec{k})|_{\phi_R}}{\phi_R(r,\o,
 \vec{k})}
\bigg)
\bigg|_{\phi_0 =0},
 \ee
 where
\be
\Pi = -\sqrt{-g} g^{rr} \del_r \phi
\ee
is the canonical momentum conjugate to $\phi$,
taking $r$ as the ``time" direction,
$\phi_R$ is a classical solution which should
 be purely in-falling at the black hole horizon and
approaches to $\phi_0(\o,\vec{k})$ asymptotically.
In order for \eq{GR} to give a
well-defined result independent of $r$,  a certain factor $r^N$ is inserted
whose power depends on the asymptotic behaviour of the metric as well as
the solution $\phi_R$. For the
standard
AdS case, it is $N={2(\D-d)}$
where $\D$ is the conformal dimension of the operator $\cO_\phi$. Finally
the subscript $\phi_0 =0$
means that one should take the part that is independent of $\phi_0$.

It is remarkable that the right hand side of (\ref{GR}) needs proper 
holographic renormalization.
Such a renormalization may affect the overall normalization of the 
two-point functions, and is well-understood in the context of 
usual AdS/CFT correspondence\cite{Skenderis:2008dg}. However, for
the warped spacetimes, which are not asymptotical to AdS, 
the holographic renormalization procedure 
has not been analyzed carefully. In this sense, 
all the results based on (\ref{GR}) should be
taken with care. Nevertheless, from the study of 
real-time correlators in the warped AdS/CFT correspondence
we have learned that the prescription (\ref{GR}) has worked 
quite well\cite{Chen:2009cg}. It will be
interesting to study the holographic renormalization 
in the warped AdS/CFT correspondence in details and confirm
the prescription \eq{GR}.

The prescription works well not just for asymptotic AdS metric, but also
for the warped AdS/CFT correspondence \cite{Chen:2009cg}.
For the Kerr/CFT correspondence,
the NHEK or the near-NHEK geometry is a warped AdS${}_3$  spacetime with the
warping factor being a function of the angular variable $\th$. Since this
modification is quite simple,
the prescription \eq{GR}
actually factorizes into a part which depends
on $\th$ only and a part which is a function of the frequencies.
Therefore one can have a well-defined prescription by taking
only the angular independent part. In general, it is the asymptotic
behaviour of
the classical solution and its conjugate momentum that matter in \eq{GR}.
For example, for a scalar field with  the asymptotic behaviour
  \be\label{asym}
 \phi\sim A(\o,\vec{k}) r^{-n_A}+B(\o,\vec{k})r^{-n_B},
 \ee
with $n_A> n_B$, the
real-time correlator of the scalar field is
given by $A(\o,\vec{k})/B(\o,\vec{k})$, up to a constant factor
independent of $\o$ and $\vec{k}$ which depends on the normalization of the
operator.
This simple result generalizes to other kinds of fields
as we will show below.

\subsection{Scalar field}

Let us consider the scalar field $\Phi$ of mass $\mu$ in the
 background (\ref{near-NHEK}). Since
there are two translational Killing vector along
 $t$ and $\phi$, we may take the ansatz:
 \be\label{ansatz}
 \Phi=e^{-i\o t+im\phi}\cR(r)\cS(\th),
 \ee
where $\o$ and $m$ are the quantum numbers. The angular part
$\cS(\th)$ satisfies the spheroidal harmonic equation:
 \be
 \frac{1}{\sin\th}\frac{d}{d\th}\left(\sin\th\frac{d}{d\th}\cS\right)+\left(
 \L_{lm}-(\frac{m^2}{4}-J\mu^2)\sin^2\th-\frac{m^2}{\sin^2\th}\right)\cS=0,
 \ee
where $\L_{lm}$ is the eigenvalue, which
can be computed
numerically. The radial part $\cR$ satisfies the equation
 \be
 \frac{d}{dr}\left(r(r+2\a)\frac{d}{dr}\right)\cR(r)-\left(
 \L_{lm}-m^2+2J\mu^2-\frac{(\o+m(r+\a))^2}{r(r+2\a)}\right)\cR(r)=0.
 \ee
Taking into account of the ingoing boundary condition at the
horizon, the radial wave function is
 \be
 \cR(r)=Nr^{-\frac{i}{2}(m+\frac{\o}{\a})}
\left(\frac{r}{2\a}+1\right)^{-\frac{i}{2}(m-\frac{\o}{\a})}
F\left(\frac{1}{2}+\b-i m, \frac{1}{2}-\b-i m,
 1-i(m+\frac{\o}{\a}); -\frac{r}{2\a}\right),
 \ee
 where
 \be
 \b^2=\frac{1}{4}+ \L_{lm}-2m^2+2J\mu^2.
 \ee
 At asymptotic infinity, the radial eigenfunction has the
 behaviour
 \be
 \cR(r)\sim Ar^{-\frac{1}{2}-\b}+Br^{-\frac{1}{2}+\b}
 \ee
 where
 \bea
 A&=&N\frac{\G(-2\b)\G(1-i(m+\frac{\o}{\a}))}
{\G(\frac{1}{2}-\b-i m)\G(\frac{1}{2}-\b-i \frac{\o}{\a})}
 (2\a)^{\frac{1}{2}+\b-\frac{i}{2}( m+\frac{\o}{\a})}, \\
 B&=& A(\b\to -\b)
 \eea
and $N$ is an arbitrary constant.
Without loss of generality, let us consider a real $\b>0$,
the prescription \eq{GR} gives the retarded correlator
 \bea \label{GR-1}
 G_R &\sim & \frac{A}{B}\nn\\
  &=&(2\a)^{2\b} \frac{\G(-2\b)}{\G(2\b)}\frac{\G(\frac{1}{2}+\b-i m)
\G(\frac{1}{2}+\b-i \frac{\o}{\a})}
  {\G(\frac{1}{2}-\b-i m)\G(\frac{1}{2}-\b-i \frac{\o}{\a})}.
 \eea
By comparing the arguments of the Gamma functions, one have the
following identification
\be \label{iden-1}
 h_L=h_R=\frac{1}{2}+\b,~~\o_L=m, ~~\bo_R=\o, ~~T_L=\frac{1}{2\pi},
 ~~T_R = T_R,
\ee
in order for \eq{GER} to be satisfied. In fact at the Matsubara
frequencies,
the expression \eq{GR-1} agrees precisely
with \eq{GE} up to an irrelevant normalization
factor which depends only on $\b$, $q_L$ and $q_R$ and can be absorbed
into the normalization of the operator.
The identification \eq{iden-1} is
the same as
the original ones suggested in  \cite{Bredberg:2009pv} where our $\o$
is $n_R$ in their notation. For imaginary $\b$, the above formula stays
the same. The complex conformal weight indicates an instability of the
AdS spacetime due to pair production \cite{Bredberg:2009pv}.

 The cross section can also be read out from the retarded correlator
 directly. It is
 \be \label{sigma-1}
 \s ={\rm Im} (G_R) =\frac{(2\a)^{2\b}}{2\b \pi
 (\G(2\b))^2}\sinh(\pi(m+\frac{\o}{\a}))|\G(\frac{1}{2}+\b-i
m)\G(\frac{1}{2}+\b-i \frac{\o}{\a})|^2.
 \ee
This agree, up to an irrelevant normalization factor, with \eq{g2} as
it should be.

Finally, one can obtain the quasi-normal modes frequencies from
the poles of the retarded Green's function. In this case, they are
 \bea
 \bo_L&=&-i2\pi T_L(n_L+h_L) \nn\\
 \bo_R&=&-i2\pi T_R(n_R+h_R)
 \eea
 with $n_L,n_R$ being non-negative integers. The left part is not
 actually the quasi-normal modes since it is related to the
 quantum number of rotation. The right part
gives the
 contribution.
As a result, we obtain the
 complete frequencies of the quasi-normal modes:
 \be
\o_R=m\O_H-i2\pi T_R(n_R+h_R).
 \ee
 This has also been obtained in  \cite{Hod:2008zz} for the near-extreme Kerr
 black holes.

\subsection{Other perturbations}

 To study various kinds perturbations about near-NHEK, including
 vector, spinor and gravitational ones, we will use
 Newman-Penrose  formalism \cite{Newman:1961qr}. For simplicity, we
 focus on
the massless perturbations.
 The NP null tetrad of near-NHEK is $e^\mu_a=(l^\mu, n^\mu, m^\mu,
 m^{\ast \mu})$,
where in coordinate basis
 \bea
 l^\mu &=&\frac{1}{r(r+2\a)}(1,r(r+2\a),0, -(r+\a)),\nn\\
 n^\mu &=&\frac{1}{4J\G(\th)}(1,-r(r+2\a),0,-(r+\a)),\nn\\
 m^\mu &=&\frac{1}{2\sqrt{J\G(\th)}}(0,0,1,i\L^{-1}(\th)),
 \eea
satisfy the normalization and orthogonal condition with
nonvanishing inner products
 \be
 l\cdot n=-m\cdot m^\ast =-1.
 \ee

 It turns out that the equations of motions of the
 perturbations
can be decomposed into two separated equations of
 motions. The wave function could be decomposed into the form
  \be
  \Psi^s=e^{-i\o t+im\phi}(r(r+2\a))^{-s}\cR^s(r)\cS^s(\th).
  \ee
 Here the angular function $\cS^s(\th)$ obeys the equation
    \be \label{angular}
 \frac{1}{\sin\th}\frac{d}{d\th}\left(\sin\th\frac{d}{d\th}\cS^s(\th)\right)
+\left(
 \L^s_{lm}-\frac{m^2}{4}\sin^2\th-ms\cos\th
-\frac{(m+s\cos\th)^2}{\sin^2\th}\right)\cS^s(\th)=0,
 \ee
where $ \L^s_{lm}$ is the separation parameter, depending on
 $l,m,s$ and satisfying
 \be\label{Lam}
 \L^s_{lm}(s)=\L^s_{lm}(-s).
 \ee

The radial function obeys
 \be
 \frac{d}{dr}\left(r(r+2\a)\frac{d}{dr}\right)\cR^s(r)-\left(
 \L^s_{lm}+q^2-2m^2-\frac{(\o+q(r+\a))^2}{r(r+2\a)}\right)\cR^s(r)=0.
 \ee
 where
 \be
 q=m-is.
 \ee
The ingoing solution is \be
 \cR^s(r)=Nr^{-\frac{i}{2}(q+\frac{\o}{\a})}
\left(\frac{r}{2\a}+1\right)^{-\frac{i}{2}(q-\frac{\o}{\a})}
 F\left(\frac{1}{2}+\b-i q, \frac{1}{2}-\b-i q,
 1-i(q+\frac{\o}{\a}); -\frac{r}{2\a}\right),
 \ee
 where
 \be\label{beta}
 \b^2=\frac{1}{4}+ \L^s_{lm}-2m^2.
 \ee
The asymptotic behaviour of the solution is
 \be
 \cR^s(r)\sim A^s r^{-\frac{1}{2}-\b} + B^s r^{-\frac{1}{2}+\b},
\ee
 where
 \bea
 A^s&=&N\frac{\G(-2\b) \G(1-i(q+ \frac{\o}{\a}))}
{\G(\frac{1}{2}-\b-iq)\G(\frac{1}{2}-\b-i\frac{\o}{\a})}
(2\a)^{\frac{1}{2}+\b-\frac{i}{2}(q+\frac{\o}{\a})}\\
B^s&=&A^s(\b \to -\b)
 \eea

 Naively one may be tempted to take the retarded Green's function to be
 proportional to $\frac{A^s}{B^s}$, as in scalar case. This is not
 true. In the usual AdS/CFT correspondence, the prescription to
 get the retarded Green's function is (\ref{GR}).
 In our case,
 things
are more subtle. In fact, for $|s|=1,2$, $\Psi^s$ are
related to the
 gauge field strength and the Weyl tensor of the tensor field:
 \bea
 \Psi^1&=& F_{\mu\nu}l^\mu m^\nu \nn\\
 \Psi^{-1}&=&(1-i\cos\th)^2F_{\mu\nu} m^{\ast\mu}n^\nu \nn\\
 \Psi^2&=& C_{\mu\nu\rho\sigma}l^\mu m^\nu l^\rho m^\sigma \nn\\
 \Psi^{-2}&=& (1-i\cos\th)^4  C_{\mu\nu\rho\sigma}n^\mu m^{\ast \nu} n^\rho m^{\ast \sigma }
 \eea
Therefore it is not  appropriate to identify $\Psi^s$ as the perturbations
 themselves. Nevertheless,  we can inversely obtain the vector and gravitational perturbations from the wave functions \cite{Chrzanowski:1975wv} in terms of the Newman-Penrose
 complex spin coefficients 
 \bea
 A_\mu &=&-(-l_\mu(\d^\ast+2\b^\ast+\t^\ast)+m^\ast_\mu(D+\r^\ast))\frac{2}{B_1}(r(r+2\a))R^{-1}(r)\cS^1(\th)e^{-i\o t+im\phi}, \\
 h_{\mu\nu}&=&\left\{-l_\mu l_\nu(\d^\ast+\l+3\b^\ast-\t^\ast)(\d^\ast+4\b^\ast+3\t^\ast)-m^\ast_\m m^\ast_\n(D-\r^\ast)(D+3\r^\ast)\right. \nn\\
  & & \left. +l_{(\m}m^\ast_{\n)}\left((D+\r-\r^\ast)(\d^\ast+4\b^\ast+3\t^\ast)+(\d^\ast+3\b^\ast-\l-\pi-\t^\ast)(D+3\r^\ast)\right)\right\} \nn\\
  & & \times \frac{4}{B_2}(r(r+2\a))^2R^{-2}(r)\cS^2(\th)e^{-i\o t+im\phi},
  \eea
  where $B_1$ and $B_2$ are two normalization factors depending on $\L_{lm}$ and $m$. In the above relations,  the differential operators $D$ and $\d^\ast$ are defined as
  \be
  D=l^\m \p_\m, \hs{5ex} \d^\ast =m^{\ast \m}\p_\m,
  \ee
  and the spin coefficients are 
  \bea
  \r =-\frac{1}{1-i\cos\th}, & \hs{3ex}& \b=\frac{1}{2\sqrt{2J}}\frac{\cos\th}{(1+i\cos\th)\sin\th}, \nn\\
  \t =-\frac{1}{\sqrt{2J}}\frac{i\sin\th}{1+\cos^2\th}, &\hs{3ex} & \l=\frac{1}{2\sqrt{2J}}\frac{-\cos\th+i(1+\sin^2\th)}{(1-i\cos\th)^2\sin\th},\nn\\
  \pi=\frac{1}{\sqrt{2J}}\frac{i\sin\th}{(1-i\cos\th)^2}. & \hs{3ex}& 
  \eea
  It is straightforward but tedious to compute the perturbations from 
the wave eigenfunctions $\Psi^s$.
 
 Note that in determining the retarded
 Green's function from gravity, it is the asymptotic behaviours of
 the source and the response that matter. In other words, once the
 source term of the field is decided, its field strength has the
 same Gamma function dependence, up to a factor. Even though
the $\Psi^s$'s are
related to the field strength, the
 source term is still proportional to $B^s$, at most up to a
 factor which plays no essential role. Once the relative
 normalization between $A^s$ and $B^s$ is fixed, it is safe to
 take $B^s$ as the source. 

 There is another tricky point. Actually the response to the
 source $B^s$
cannot be taken as the $A^s$ term in the same $\Psi^s$.
Instead, the response
should be given by
the $A^{-s}$ term in  $\Psi^{-s}$.
For example, for the fermionic perturbations, it has been shown
 that the conjugate momentum to $\psi^-$ is proportional to
 $\psi^+$ \cite{Iqbal:2009fd}.  And in the study of warped AdS/CFT
 correspondence, it was found that for vector perturbation, the
 conjugate momentum of $A_i$ is not in itself, but in another
 component of $A_\mu$ \cite{Chen:2009cg}. In fact, for the gauge
 field,
the conjugate momentum
 of $A_\mu$ is the field strength $F^{r\mu}$, which is related
to $\Psi^s$ directly.
 Here, even if we have no rigorous derivation, taking into account of
 above two points we would like to propose the following
 working prescription for computing the retarded Green's function
for general perturbations with spin
 $s$:
 \be\label{pres}
 G_R^s \sim (-1)^s\frac{A^{-s}}{B^{s}}.
 \ee
 Here $(-1)^s$ is mainly for the fermionic perturbations.

 With this prescription,  the retarded
 Green's function of the perturbations with spin $s$ is
 given by
 \bea
 G_R^s &\sim& (-1)^s\frac{A^{-s}}{B^{s}} \nn\\
   &\sim & (-1)^s(2\a)^{2\b}\frac{\G(-2\b)\G(\frac{1}{2}+\b-s-i m)
\G(\frac{1}{2}+\b-i\frac{\o}{\a})}
 {\G(2\b)\G(\frac{1}{2}-\b+s-i
 m)\G(\frac{1}{2}-\b-i\frac{\o}{\a})}.
 \eea
To get the second line, we have used the relation (\ref{Lam}).
With the conformal weights of the field identified as
 \be
 h^s_R=\frac{1}{2}+\b,\hs{5ex} h^s_L=h^s_R-s,
 \ee
 the retarded Green's function can be rewritten as
 \be
 G^s_R\sim (-1)^sT_R^{2h^s_R-1}\frac{\G(1-2h^s_R)\G(h^s_L-i m)
\G(h^s_R-i\frac{\o}{\a})}
 {\G(2h^s_R-1)\G(1-h^s_L-i
 m)\G(1-h^s_R-i\frac{\o}{\a})}.
 \ee
It is straightforward to check that at the Matsubara
frequencies,  the above retarded Green's
function agrees precisely,  up to
a frequencies independent normalization factor, with the CFT result
\eq{GE} if the frequencies and the temperatures are identified as before:
\be
\o_L=m, \quad \o_R =\o,\quad T_L =\frac{1}{2\pi}, \quad T_R=T_R.
\ee

The cross section can be read
 directly from the Green's function by the relation $\s \sim {\rm Im}
 (G_R)$.
It turns out that for the fermion, the cross section is
 \be
 \s \sim
 \frac{(2\a)^{2h_R-1}}{\G(2h_R-1)^2}\cosh((m+\frac{\o}{\a})\pi)|\G(h_L-i
 m)\G(h_R-i\frac{\o}{\a})|^2.
 \ee
And for the gauge field and the graviton, the cross sections are
in the same form as
 \be
    \s \sim
 \frac{(2\a)^{2h_R-1}}{\G(2h_R-1)^2}\sinh((m+\frac{\o}{\a})\pi)|\G(h_L-i
 m)\G(h_R-i\frac{\o}{\a})|^2.
 \ee
They agree with the CFT result.

As for the quasi-normal modes, their frequencies are simply
 \bea
 \o^s_L &=& -i2\pi T_L (h^s_L+n_L) \nn\\
 \o^s_R &=& m\O_H -i2\pi T_R (h^s_R+n_R)
 \eea
 with $n_L,n_R$ being non-negative integers.

\section{Real-time Correlators in NHEK}

It is interesting to generalize the
discussion in the last section to the NHEK geometry. At first
look, the NHEK geometry is like the vacuum with
the near-NHEK geometry as excitation.
However,  since NHEK itself is dual to the 2D
chiral CFT with a temperature in the left-moving sector,
it is thus more
natural to take NHEK as a
limiting case of near-NHEK.

To discuss the perturbations scattering in NHEK geometry, it is
not appropriate to take the $\a \to 0$ limit since it leads to singularity
in the eigenfunctions. To simplify the discussion, let us consider the
massless perturbations so that we can treat all kinds of
perturbations at the same time. The NP null tetrad could be
taken as the $\a \to 0$
limit of the one in near-NHEK. We start with the same ansatz
 \be
 \Phi^s =e^{-i\o t+im\phi}(r)^{-2s}\cR^s(r)\cS^s(\th).
 \ee
The angular function $\cS^s(\th)$ satisfies the same equation
(\ref{angular}), while the radial  function obeys
 \be
 \frac{d}{dr}\left(r^2\frac{d}{dr}\right)\cR^s(r)-\left(
 \L_{lm}+q^2-2m^2-\frac{(\o+qr)^2}{r^2}\right)\cR^s(r)=0.
 \ee
 where
 \be
 q=m-is.
 \ee
 Introduce $z=1/r$, we can rewrite this equation into
 \be
 \frac{d^2}{dz^2}\cR^s+\left(\frac{\frac{1}{4}-\b^2}{z^2}
+\frac{2q\o}{z}-\o^2\right)\cR^s=0,
 \ee
 where $\b$ is the same as defined in (\ref{beta}).
 The solution could be written in terms of Kummer function
 \be
 \cR^s_\pm=e^{-i\o z}(2i\o z)^{\frac{1}{2}\pm \b}F(\frac{1}{2}\pm \b
 -iq, 1\pm 2\b;2\o z).
 \ee
 We take the point of view that the horizon is at $r=0$, and
 requires the ingoing boundary condition at the horizon. Then we
 find that the eigenfunction is the combination of the above two
 functions:
 \be
 \cR^s=A^s_+\cR^s_+ + A^s_- \cR^s_-,
 \ee
 where
  \be
  A^s_+=-\frac{\G(1-2\b)}{\G(\frac{1}{2}-\b-iq)}A_0,\hs{5ex}
  A^s_-=\frac{\G(1+2\b)}{\G(\frac{1}{2}+\b-iq)}A_0,
 \ee
 with $A_0$ being a constant.

 Note that our discussion on the perturbations is different from the
one in \cite{dias}, in which the treatment of NHEK geometry was in
global coordinate. We would like to emphasize that the coordinate
transformation from NHEK to global coordinate is singular. It is
essential to take NHEK geometry as an extreme black hole such that
its correspondence to 2D CFT at finite temperature is transparent.

 Following the prescription, we have the retarded Green's function
  \bea
  G^s_R &\sim &  (-1)^s\frac{A^{-s}_+}{A^s_-} \nn\\
   &=& (-1)^s\frac{\G(-2\b)\G(\frac{1}{2}+\b-s-i m)}{\G(2\b)\G(\frac{1}{2}-\b+s-i
   m)}\\
   &=& (-1)^s\frac{\G(1-2h_R)\G(h_L-i m)}{\G(2h_R -1)\G(1-h_L-i
   m)}
   \eea
   where we have identified
   \be
   h_R=\frac{1}{2}+\b, \hs{5ex} h_L=h_R-s.
   \ee
It is easy to see that the Green's functions are in consistence
with the prediction of a  CFT with only left-moving
modes of frequency
\be
\o_L =m.
\ee
Therefore the NHEK background looks like an
extreme black hole, corresponding to a ``chiral" CFT at $T_L =
1/2\pi$  \cite{Chen:2009cg}.
The cross sections follow from the retarded Green's function. For
the fermionic perturbation,
we have
 \be
 \s \sim \frac{\G(2(1-h_R))}{\G(2h_R)}\cos(h_R\pi)\cosh(m\pi)|\G(h_L-i
 m)|^2,
 \ee
while for the bosonic perturbation,
we have
\be\label{NHEKsec}
 \s \sim \frac{\G(2(1-h_R))}{\G(2h_R)}\cos(h_R\pi) \sinh(m\pi)|\G(h_L-i
 m)|^2.
 \ee
They are in agreement with the CFT results.
The retarded correlation has the simple poles at
 \be
 m=-i2\pi T_L (n_L+h_L).
 \ee
 Note that now the poles have nothing to do with the frequencies,
 so there is no quasi-normal mode for extreme Kerr black hole.

\section{Kerr-Newman Case}

For the Kerr-Newman black hole with mass $M$, angular momentum $J=aM$
and
electric charge $Q$, its metric takes the following form
\be
ds^2=-\frac{\D}{\r^2}(d\hat t-a\sin^2\th d\hat \phi)^2+
\frac{\r^2}{\D}d\hat r^2+\rho^2 d\th^2+
\frac{1}{\r^2}\sin^2\th\left(ad\hat t-(\hat r^2+a^2)d\hat \phi\right)^2,
\ee
where
\bea
\Delta&=&(\hat r^2+a^2)-2M\hat r+Q^2, \nn\\
\r^2&=&\hat r^2+a^2\cos^2\th.
\eea
There are two horizons located at
\be
\hat r_\pm=M\pm \sqrt{M^2-a^2-Q^2}.
\ee
And the Hawking temperature, entropy, angular velocity of the
horizon and the electric potential are respectively
 \bea
 T_H&=&\frac{\hat r_+-\hat r_-}{4\pi(\hat r_+^2+a^2)},\nn\\
 S&=&\pi(\hat r_+^2+a^2),\nn\\
 \O_H&=&\frac{a}{\hat r_+^2+a^2},\nn\\
 \Phi&=&\frac{Q\hat r_+}{\hat r_+^2+a^2}.
 \eea

For the extreme case, $\hat r_+=\hat r_-$ such that $T_H=0$.
However, from the first law of thermodynamics, we have
nonvanishing left-moving temperature and left-moving chemical potential
\be
T_L=\frac{\hat r^2_++a^2}{4\pi J}, \hs{5ex}\mu_L=-\frac{Q^3}{2J}.
\ee

The near horizon geometry of extreme Kerr-Newman (NHEK-Newman) could
be
obtained by the same
scaling limit as the one in extreme Kerr geometry
\cite{Bardeen:1999px,{Hartman:2008pb}}. It is given by
\be
ds^2=\G(\th)\left(-r^2dt^2+\frac{dr^2}{r^2}+d\th^2\right)+\L(\th)(d\phi+brdt)^2,
\ee
where
\bea
\G(\th)&=&\hat r^2_++a^2\cos^2\th \nn\\
\L(\th)&=&\frac{(\hat r_+^2+a^2)^2\,sin^2\th}{\hat r^2_++a^2\cos^2\th},\nn\\
b&=&\frac{2a\hat r_+}{\hat r^2_++a^2}.
\eea
The gauge potential and the field strength are
\bea
A&=&\frac{Q}{b}\frac{\hat r^2_+-a^2\cos^2\th}{\hat r^2_++a^2\cos^2\th}
(d\phi+br dt),\\
F&=&-Q\frac{\hat r^2_+-a^2\cos^2\th}{\hat r^2_++a^2\cos^2\th}(d\phi+br
dt)dt
\wedge dr+Q\frac{(\hat r_+^2+a^2)\hat R_+ a\sin\th\cos\th}
{(\hat r^2_++a^2\cos^2\th)^2}d\th \wedge(d\phi+brdt).\nn
\eea

For the near-extremal Kerr-Newman black hole, the modes in the
right-moving
sector are excited. Taking a scaling
limit with a finite right-moving temperature, 
we obtain the near-NHEK-Newman geometry
\be
ds^2=\G(\th)\left(-r(r+2\a)dt^2+\frac{dr^2}{r(r+2\a)}+d\th^2\right)
+\L(\th)(d\phi+b(r+\a)dt)^2,
\ee
with $\a=2\pi T_R$. Similar to the Kerr black hole case, the
near-NHEK-Newman geometry looks like a black hole with horizon at $r=0$,
while NHEK-Newman geometry is an extreme black hole with the horizon at $r=0$.

For simplicity, we will focus on the complex \
scalar field with mass $\mu$ and charge $e$. The Klein-Gordon equation is
\be
(\nabla_\mu+ieA_\mu)(\nabla^\mu+ieA^\mu)\Phi-\mu^2\Phi=0.
\ee
With the ansatz (\ref{ansatz}), the angular wave function satisfies
\be
 \frac{1}{\sin\th}\frac{d}{d\th}\left(\sin\th\frac{d}{d\th}\cS\right)+\left(
 \L_{lm}-a^2(\o^2_0-\mu^2)\sin^2\th-\frac{m^2}{\sin^2\th}\right)\cS=0.
 \ee
Here $\L_{lm}$ is the separation constant. It is
restricted by the regularity
boundary condition at $\th=0,\pi$ and can be computed
numerically. $\o_0$ is the frequency saturate the superradiant bound
\be
\o_0=m\O_H+e\Phi.
\ee

The radial part $\cR$ satisfies the equation
 \be
 \frac{d}{dr}\left(r(r+2\a)\frac{d}{dr}\right)\cR(r)-\left(
 \L_{lm}+\mu^2(\hat r_+^2+a^2)-2am\o_0-
\frac{(\o+\tilde m(r+\a))^2}{r(r+2\a)}\right)\cR(r)=0,
 \ee
where
\bea
\tilde m&=&bm+\frac{\hat r_+^2-a^2}{\hat r_+^2+a^2}eQ \nn\\
 &=&\frac{m-e\mu_L}{2\pi T_L}
\eea
The radial eigenfunction which is ingoing at the horizon $r=0$ is
\be
\cR(r)=r^{-\frac{i}{2}(\tilde m+\frac{\o}{\a})}\left(\frac{r}{2\a}
+1\right)^{\frac{i}{2}(\frac{\o}{\a}-\tilde m)}F(\frac{1}{2}
+\b-i\tilde m,\frac{1}{2}-\b-i\tilde m,1-i(\tilde m+\frac{\o}{\a})
;-\frac{r}{2\a}),
\ee
where
\be
\b^2 =\frac{1}{4}+\L_{lm}-2am\o_0-\tilde m^2+\mu^2(M^2+a^2).
\ee
Asymptotically, the radial eigenfunction could be expanded as
\be
\cR(r)\sim Ar^{-\frac{1}{2}-\b}+Br^{-\frac{1}{2}+\b},
\ee
where
\bea
A&=&N\frac{\G(-2\b)\G(1-i(\tilde m+\frac{\o}{\a})}
{\G(\frac{1}{2}-\b-i\tilde m)\G( \frac{1}{2}-\b-i\frac{\o}{\a})}
(2\a)^{\frac{1}{2}+\b-\frac{i}{2}(\tilde m+\frac{\o}{\a})}\\
B&=&A(\b \to -\b),
\eea
with $N$ being a constant. Compared to the near-NHEK case,
there are two differences. One is the value of $\b$. However the
conformal weight of the scalar is still
\be
h_R=\frac{1}{2}\pm \b,
\ee
so this difference does not affect the discussion. The other
difference is on the angular momentum dependence. Now we have $\tilde
m$,
which has been shifted by the
chemical potential in left-moving sector.

The retarded Green's function is
\bea
G_R &\sim& \frac{A}{B} \nn\\
    &=& \frac{\G(-2\b)}{\G(2\b)}\frac{\G(\frac{1}{2}+\b-i \tilde m)
\G(\frac{1}{2}+\b-i \frac{\o}{\a})}
  {\G(\frac{1}{2}-\b-i \tilde m)\G(\frac{1}{2}-\b-i \frac{\o}{\a})}(2\a)^{2\b}.
 \eea
Taking the following identification into account,
\be
h_L=h_R=\frac{1}{2}+\b, \hs{3ex}
\o_L= m - e \mu_L , \hs{3ex}
\bo_R=\o, \hs{3ex}
T_L=\frac{M^2+a^2}{4\pi M a},\hs{3ex}T_R=T_R,
\ee
the retarded Green's function and the cross section are both in good
agreement with the CFT prediction.
Note that due to the coupling with the background electric field, the
angular quantum number $m$ gets shifted.
In other words, the presence of the electric field affect only the
left-moving sector.

Similarly, the quasi-normal modes could be read from the poles in
the retarded Green's function. These poles are located at
\bea
\tilde m&=&-i(n_L+h_L)\nn\\
\o&=&-i2\pi T_R(n_R+h_R),
\eea
with $n_L,n_R$ being non-negative integers.
The second relation gives the quasi-normal modes
\be
\o_R=m\O_H-i2\pi T_R(n_R+h_R),
\ee
after taking into account of the superradiant bound.

For the case of NHEK-Newman geometry, the analysis is very similar to
the NHEK case. The radial eigenfunction
of complex massive scalar could be written in terms of Kummer
function. And the
retarded Green's function and the cross section
are of the same form as
\eq{GR-1} and \eq{sigma-1}
with $m$ being
replaced by $\tilde m$. This suggests that the NHEK-Newman
geometry is dual to a chiral part of 2D CFT with a chemical
potential
and a non-vanishing
left temperature $T_L$.

\section{Conclusions and Discussions}

In this paper, we studied the real-time correlators in Kerr/CFT
correspondence. Kerr/CFT correspondence states that a quantum gravity
theory in NHEK geometry is dual to chiral (left) part of a 2D CFT with
left-moving temperature. For near-extremal Kerr black holes,  there
are excited right-moving modes as well, which is dual to the
right-moving excitations in dual 2D CFT. Remarkably, this phenomenon
shows up nicely in a geometrical way, fitting well with the warped
AdS/CFT correspondence. In fact, NHEK geometry is an extreme warped
AdS$_3$ black hole for fixed $\th$, whose dual CFT has only
non-vanishing left-moving temperature; while near-NHEK geometry is a
non-extremal AdS$_3$ black hole for fixed $\th$, whose dual CFT has
both non-vanishing left- and right-moving temperatures.
Such a picture has been discussed in the warped
AdS$_3$/CFT correspondence in \cite{Chen:2009cg}.
Using the fact that the NHEK and the near-NHEK geometry can be considered
as a warped AdS geometry,
we applied the holographic
prescription in AdS/CFT correspondence and computed the retarded
correlators,
the cross sections of various perturbations
and the quasi-normal modes  in these geometries.
Perfect agreement with the CFT predictions was found.
We have also generalized the discussion to the cases of  extreme
Kerr-Newman and the near extreme Kerr-Newman.
The picture is similar and we found a new effect in the
shift of the frequency due to
the existence of an electric field.

Some of the results in this paper, including the cross-sections of
various perturbations in near-NHEK, near-NHEK-Newman have been
obtained in  \cite{Bredberg:2009pv,Hartman:2009nz}. Here we have
presented
a different derivation using the retarded Green's functions.
By considering the Kerr/CFT correspondence
as a  warped AdS/CFT correspondence, we
emphasize the possibility and necessity to consider the perturbations
entirely in near-horizon geometries.
In particular, we investigated the perturbations in
NHEK and NHEK-Newman geometries and showed that they are extreme
black holes dual to only the left-parts of 2D CFTs.
One interesting
question we would like to ask is what corresponds to a 2D CFT without
temperature.

Our treatment could also be applied to the multi-charged and
higher-dimensional black holes as discussed in
\cite{Cvetic:2009jn}.
Even though we are going
to work in higher dimension, the extra isometries will simplify the
analysis and allow us to study two decomposed master equations, similar
to what we have met. The key point is still that the master radial
equation suggests that the perturbation is scattering off a warped
black holes.
We will leave the details.

Despite many successful checks of the Kerr/CFT correspondence,
the subject is far
from understood and its consistency has been challenged, particularly
concerning the GHSS boundary conditions \cite{AndyWei}.
Recently other boundary conditions which enhance the
$SL(2,R)_R$ isometry to Virasoro in addition to
the left-moving  Virasoro algebra have been proposed
\cite{matsuo,ras}.  Due to their specific form, it is not clear
whether these generators actually
describe a symmetry or just gauge transformation \cite{marolf-st}.
The dynamics of the near horizon geometry subjected to the GHSS
boundary conditions has been examined \cite{dias,amsel}
and it has been argued that the NHEK geometry contains no dynamics.
It is important to understand better these issues.

\section*{Acknowledgments}

 BC would like to thank Grey college, Durham University for
hospitality during his visit.
 The work of BC was partially supported by NSFC Grant
No.10775002,10975005, and NKBRPC (No. 2006CB805905).
The work of CSC has been supported by STFC and EPSRC.

\end{document}